\documentclass[review]{elsarticle}

\usepackage{hyperref}

\journal{Elsevier}

\begin{document}

\begin{frontmatter}

\title{Effect of radial flow on two particle correlations with identified triggers at intermediate $p_{T}$ in p-Pb collisions at $\sqrt{s_{NN}} = $ 5.02 TeV}

\author[mymainaddress]{Debojit Sarkar\corref{mycorrespondingauthor}}
\cortext[mycorrespondingauthor]{Corresponding author}
\ead{debojit03564@gmail.com}

\author[mymainaddress]{Subikash Choudhury}
\author[mymainaddress]{Subhasis Chattopadhyay}


\address[mymainaddress]{Variable Energy Cyclotron Centre, 1/AF-Bidhannagar, Kolkata-700064, India}

\begin{abstract}
Results from the two-particle correlation between identified triggers (pions ($\pi^{\pm}$) , protons (p/$\bar{p}$))) and un-identified charged
particles at intermediate transverse momentum ($p_{T}$) in p-Pb collisions at $\sqrt{s_{NN}} = $ 5.02 TeV  have been presented.
The events generated
from a hybrid Monte-Carlo event generator, EPOS 3.107 that implements a flux-tube
initial conditions followed by event by event 3+1D viscous hydrodynamical evolution, have been analysed to calculate two-dimensional correlation functions
in $\Delta\eta$-$\Delta\phi$.
The strength of angular correlations at small relative angles (jet-like correlations), quantified in terms of near-side jet-like
per-trigger yield
has been calculated as a function of the event multiplicity. The yield associated with pion triggers exhibit negligible
multiplicity dependence, while the proton-triggered yield shows a gradual suppression from low to high multiplicity
events. In small collision systems like p-Pb where jet modification is expected to be less dominant, the observed suppression
may be associated with the hydrodynamical evolution of the bulk system
that generates an outward radial flow. Analogous measurements in Au-Au collisions at RHIC energy have shown a hint of dilution in proton-triggered
correlation at its highest multiplicity suggesting
that the soft physics processes in p-Pb and heavy ion collisions may have qualitative similarity.

\end{abstract}


\end{frontmatter}


\section{Introduction}

Collisions of small systems like p-Pb \cite {alice_pPb_double_ridge,pPb_mass_ordering, multiparticleCorr_pPb,p_pi_enhancement_pPb} or d-Au \cite {d-Au ridge paper} at LHC and RHIC energies have gained renewed interest following the discovery that the underlying collision dynamics
might be analogous to that of the heavy-ions. Two-particle angular correlation measurements in p-Pb \cite {alice_pPb_double_ridge} and d-Au \cite {d-Au ridge paper} collisions have revealed existence of
azimuthal correlations extended to a large pseudorapidity separation $|\Delta\eta|$, the effect popularly  known as ''ridge''. In heavy-ion collisions, such structures have been attributed to the effect of the initial anisotropy in the overlap geometry of the colliding nuclei followed by the collective emission of particles.

Several other observations, like,
 mass-ordering of the elliptic flow coefficient ($v_{2}$) of identified particles  \cite {pPb_mass_ordering}, quark-scaling of $v_{2}$ \cite {CMSv2paper} and baryon-to-meson enhancement at intermediate $p_{T}$  \cite {p_pi_enhancement_pPb} suggest
similarity between the systems formed in small and heavy-ion collisions.

However striking anomaly has been observed in the measurements sensitive to the energy loss. Absence
of significant jet-quenching or negligible modification of the hadon yield ($R_{pPb}$) \cite {RpPb} upto 20 GeV/c indicate that the medium may be  transparent to
the hard-QCD processes. It has been argued that angular-correlations in small systems are dominated by jet-like processes. However, the emergence of the near side ridge in high-multiplicity event classes of small collision systems (pp and p-Pb) \cite {CMS_pp_Ridge}  \cite {pPb_mass_ordering} still lacks un-ambiguous understanding. Several theoretical propositions have been made based on correlated emission from
 glasma flux tubes(CGC approach)  \cite {CGC_pPb}, collective flow due to hydro-dynamical effects \cite {epos_massordering_flow_pPb} \cite {epos_ridgein_pp} or incohrent parton scatterings \cite {AMPT_pp_pPb_ridge} \cite {AMPT_pPb_flow}, but no general agreement could be reached.

In this paper we have adopted the two-particle correlation technique to calculate  the near-side jet like yield  between identified triggers(2.0  $<p{_T}< $4.0 GeV/$\it{c}$) and un-identified associateds(1.0  $<p{_T}< $4.0 GeV/$\it{c}$) as a function of event multiplicity in p-Pb collisions at 5.02 TeV. $p_{T}$ ranges of trigger and associated particles are chosen in the region where the ridge structures have been prominently observed in the experimental results \cite {alice_pPb_double_ridge}. An inclusive baryon to meson enhancement has also been observed in this $p_{T}$ range in p-Pb collisions at 5.02 TeV \cite {p_pi_enhancement_pPb}, similar to that found in heavy ion collisions \cite {Au_Au_p_to_Pi}  \cite {Pb_Pb_P_to_Pi}, where it has been discussed in terms of radial flow and/or quark coalescence model of hadronization. Discussion on the coalescence model of hadronization \cite {RFPRL90_2003_14} \cite {VGPRL90_2003_15} towards baryon to meson enhancement at intermediate $p_{T}$ is beyond the scope of this paper. However a consistent explanation to such observation has also been offered by the models that employs hydrodynamics \cite {epos_radialflow_spectra_pPb}. Radial boost, generated during the hydrodynamical evolution, pushes the massive hadrons more to higher $p_{T}$ and provide a natural explanation to the enhanced baryon generation at intermediate $p_{T}$. Assuming the applicability of the hydrodynamics in high multiplicity p-Pb collisions, a larger radial flow has been suggested in high multiplicity p-Pb compared to central Pb-Pb collisions \cite {Shuryak_radialflow} and supported by the experimental observation \cite {p_pi_enhancement_pPb}. The effect of radial flow on the spectra at intermediate $p_{T}$ and a comparative study between EPOS3 and ALICE results \cite {epos_radialflow_spectra_pPb} will be discussed in the next section.
In this analysis the trigger particles are identified as pions or protons and correlated jet like yield (per trigger) is calculated as a function of event multiplicity after subtracting the flow-modulated back-ground estimated from large $|\Delta\eta|$. The anomalous baryon to meson enhancement is expected to alter the near-side jet yield associated with proton-triggers in a more significant way compared to the pion triggers.  Hadrons pushed from lower $p_{T}$ by radial flow are  expected not to have short range collimated jet-like correlation beyond the expected flow (ridge) like correlation. Thus once bulk is subtracted, the baryon-triggered jet like yield is expected to get suppressed more compared to the pion triggered one from low to high multiplicity event classes commonly referred to as "trigger dilution" \cite {star_triggerdilution} \cite {phenix_triggerdilution}.
We have performed the correlation analysis on events generated using EPOS 3.107. Full description of the model will be presented later in this text. Analysis has been performed for two-configurations - with and without hadronic-rescattering to study the effect of hadronic evolution
on the observable. The sensitivity of the near side proton and pion triggered jet like yield  
 towards the radial flow can be tested using EPOS3 event generator which explicitly contains hydrodynamical flow \cite {epos_model_descrip}. 
The paper is organised as follows. 
In the next section  we have given a brief description of EPOS 3.107 event generator. 
The method of extraction of background subtracted correlation function is discussed in section 3. 
The discussions on results are performed in section 4.

\section{The EPOS3 Model}

EPOS3 basically contains a hydrodynamical approach based on flux tube initial conditions \cite {epos_radialflow_spectra_pPb}  \cite {epos_model_descrip} \cite {parton_gribov_Regge_Th}. This is a parton based  model where the partons initially undergo multiple scatterings. Each scattering is composed of  hard elementary scattering with initial and final state linear parton emission- forming parton ladder or "pomeron". Each parton ladder has it's own saturation scale $Q_{s}^{2}$ depending on the number of connected participants and it's center of mass energy and this will separate the soft processes from hard/p-QCD ones. This formalism is referred as "Parton based Gribov Regge Theory" and explained in detail in \cite {epos_massordering_flow_pPb}  \cite {parton_gribov_Regge_Th}. In this formalism each ladder may be considered as a longitudinal colour field which can be treated as a relativistic string. After multiple scatterings the final state partonic system consistes of mainly longitudinal colour flux tubes carrying transverse momentum of the hard scattered partons in transverse direction. Depending on the energy of the string segments and local string density - the high density areas form the "core"(containing string segments more than a critical value per unit area in  given transverse slices) and the low density area form the "corona" \cite {epos_core_corona_sep}. The core part basically forms the bulk matter. The strings in the core thermalize and then undergo hydrodynamical expansion and finally hadronize.The strings in the corona hadronize by Schwinger's mechanism and basically constitutes the jet part(high $p_{T}$ particles) of the system. The hadrons formed in the final stage can undergo scattering. EPOS3 introduces a theoretical scheme which takes into account  event by event 3+1D viscous hydrodynamical evolution of the bulk matter, jets and their interactions \cite {epos_massordering_flow_pPb}. In the context of  this model , multiplicity of an event is proportional to the number of pomerons whose positions are generated randomly and may lead to an elliptical shape of the core part eventually creating a cos(2$\Delta\phi$) shaped correlation function \cite {epos_massordering_flow_pPb}. The translational invariance of the structure finally generates a double ridge. In \cite {epos_massordering_flow_pPb}, EPOS3 has reasonably reproduced the double ridge and mass ordering of $v_{2}$ of identified particles in high multiplicity p-Pb collisions as observed by the ALICE collaboration at CERN \cite {alice_pPb_double_ridge} \cite {pPb_mass_ordering}. This  model has also successfully explained the nuclear modification factor $R_{pPb}$ for NSD (non single-diffractive) p-Pb collisions at $\sqrt{s_{NN}}$ = 5.02 TeV as measured by ALICE \cite {epos_radialflow_spectra_pPb} \cite {RpPb} .

 In section VIII of \cite {epos_radialflow_spectra_pPb}, the multiplicity dependence of the identified particle spectra (as measured by ALICE in p-Pb collisions at 5.02 TeV) has been compared with different event generators (e.g. QGSJETII, AMPT and EPOS3).  Models like QGSJETII(contains  no flow) and AMPT (contains little flow) underpredicts the spectra at intermediate $p_{T}$ (2.0  $<p{_T}< $4.0 GeV/$\it{c}$). The mismatch with the data is more prominent for heavier particles (protons, lambdas) compared to the low mass ones(e.g. pions) indicating the need of flow like effect(mainly radial flow) to explain the spectra at intermediate $p_{T}$. The hydrodynamic flow employed in EPOS3 helps to describe the spectra at intermediate $p_{T}$. The radial flow pushes heavier particles more compared to the lighter ones from lower to higher $p_{T}$  values describing the scenario at intermediate $p_{T}$ in a better way. Furthermore the multiplicity dependence of the particle ratios measured by ALICE has been compared with the output of different event generators mentioned above \cite {epos_radialflow_spectra_pPb}. The trend of baryon to meson enhancement at intermediate $p_{T}$ has been observed in EPOS3 \cite {epos_radialflow_spectra_pPb}.

\begin{figure}
\begin{center}
\includegraphics[height=7.0 cm, width=11.0 cm]{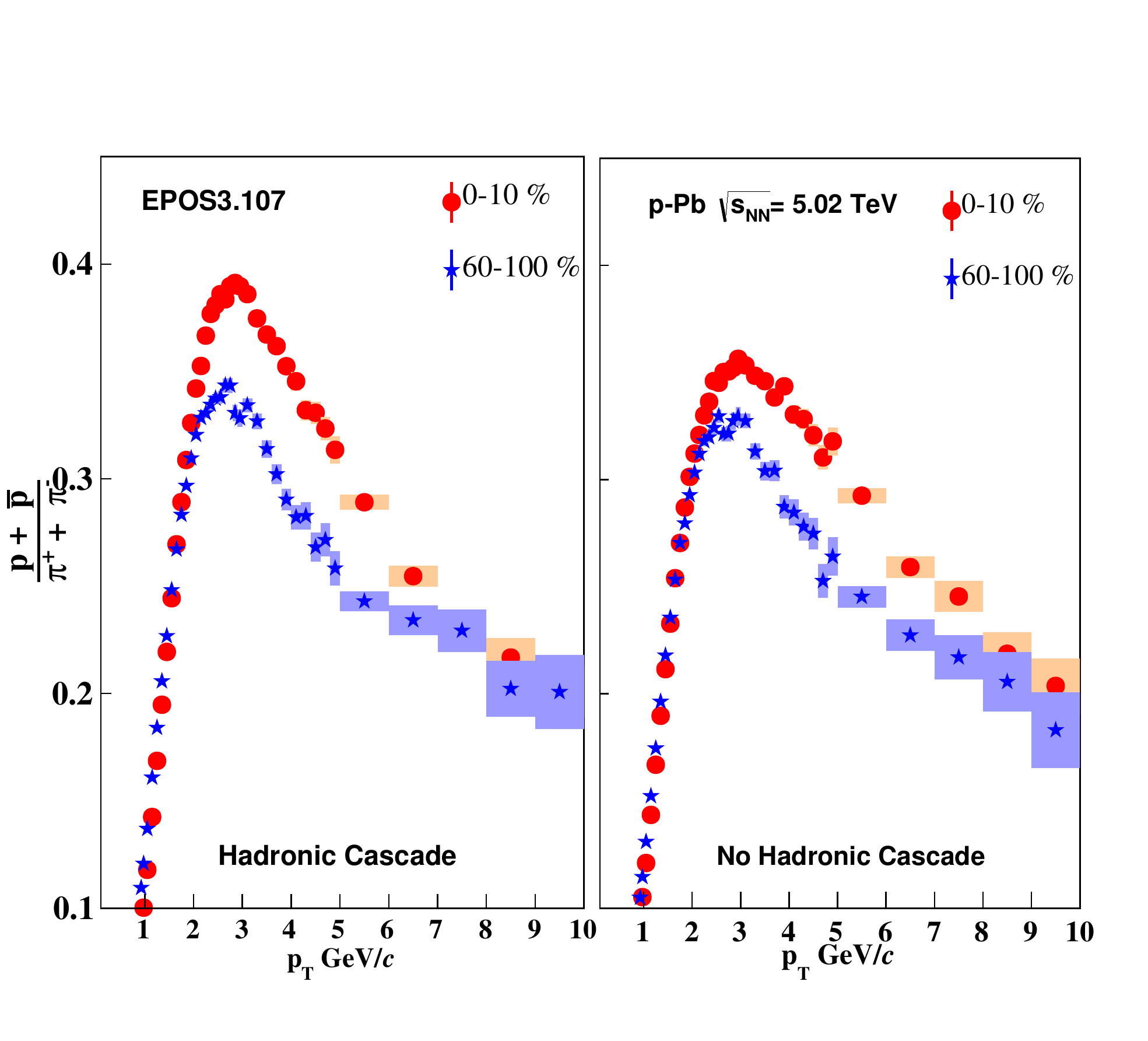}
\caption{[Color online] Inclusive proton over pion ratio as obtained from EPOS 3.107
in 0-10\% and 60-100\% event classes with(left) and without(right) hadronic cascade.}
\label{inclusive-ratio}
\end{center}
\end{figure}

In Fig.~\ref{inclusive-ratio}. the inclusive proton to pion ratio as obtained from EPOS 3.107 is given with hadronic scattering  on and off. In both cases a clear p/$\pi$ enhancement at intermediate $p_{T}$ has been observed. The hadronic rescattering shows a slight enhancement in the ratio indicating the generation of mild radial flow during hadronic evolution.
In this paper, we have performed two particle correlation analysis with trigger particles from intermediate $p_{T}$ (2.0  $<p{_T}< $4.0 GeV/$\it{c}$). The goal of this  study is to evaluate how radial flow affects the jet like per trigger yield associated with proton and pion triggers in different multiplicity classes.

\section{Analysis Method}

Two particle correlation is a widely used technique in high-energy physics for extracting the properties of the system 
produced in the ultra relativistically high energetic collisions. In the present analysis the correlation function is obtained 
among two sets of particles classified as {\it trigger} and {\it associated}. The p${_T}$ range of trigger and 
associated particles are 2.0  $<p{_T}<$ 4.0 GeV/$\it{c}$ and 1.0 $<p{_T}<$ 4.0 GeV/$\it{c}$ respectively and the correlation function has been constructed with a $p_{T}$ ordering ($p_{T}^{assoc} < p_{T}^{Trigger}$). The pseudo-rapidity of the particles are 
restricted within -0.8$<\eta<$0.8.  A 2D correlation function is obtained as a function of the difference in 
azimuthal angle $\Delta\phi$ = $\bf\phi_{trigger}$ -$\bf\phi_{associated}$ and pseudo-rapidity $\Delta\eta$
= $\bf\eta_{trigger}$ -$\bf\eta_{associated}$. The same event correlation function is defined as 
$\frac{dN_{same}}{N_{trigger}d\Delta\eta d\Delta\phi}$, where $N_{same}$ is the number of particles associated to 
triggers particles ($N_{trigger}$) that are taken from the same event. To correct for pair acceptance the same event correlation function is divided by mixed event correlation function  $\alpha$$\frac{dN_{mixed}}{d\Delta\eta d\Delta\phi}$. The mixed event correlation function is constructed by correlating the trigger particles in one event to the associated particles from other events belonging to the same multiplicity event class. The factor $\alpha$ is used to normalize  the mixed event to make it unity for pairs where both particles go into approximately the same direction ($|\Delta\eta|$ $\approx$ 0, $|\Delta\phi|$ $\approx$ 0).

Both p/$\pi$ ratio and correlation analysis have been performed by dividing the entire minimum bias events
into four multiplicity classes based on the total number of charged particles produced (with $p_{T}$ $>$0.05 GeV/c) within 2.8  $<\eta<$5.1. This is the acceptance range of ALICE VZERO-A detector in the Pb going direction in case of p-Pb collisions and used for multiplicity class determination by the ALICE collaboration \cite {alice_pPb_double_ridge},\cite {pPb_mass_ordering}. The multiplicity classes are denoted as 60-100$\%$, 40-60$\%$, 10-40$\%$, 0-10$\%$ from the lowest to the highest multiplicity.
Two particle correlation functions with proton and pion triggers in the 0-10$\%$ event class are given in  Fig.2.

\begin{figure}[htb!]
\includegraphics[scale=0.37]{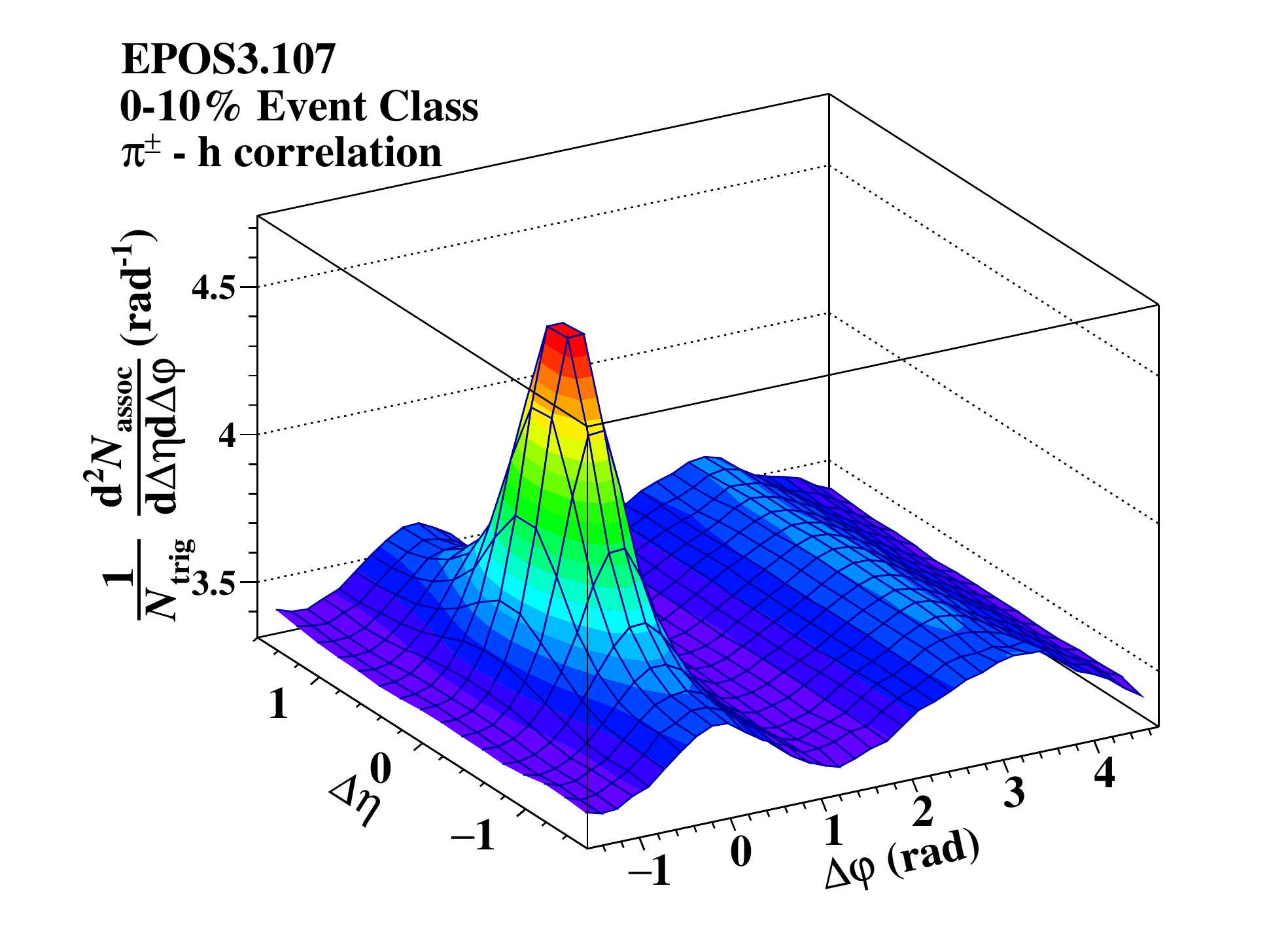}
\includegraphics[scale=0.36]{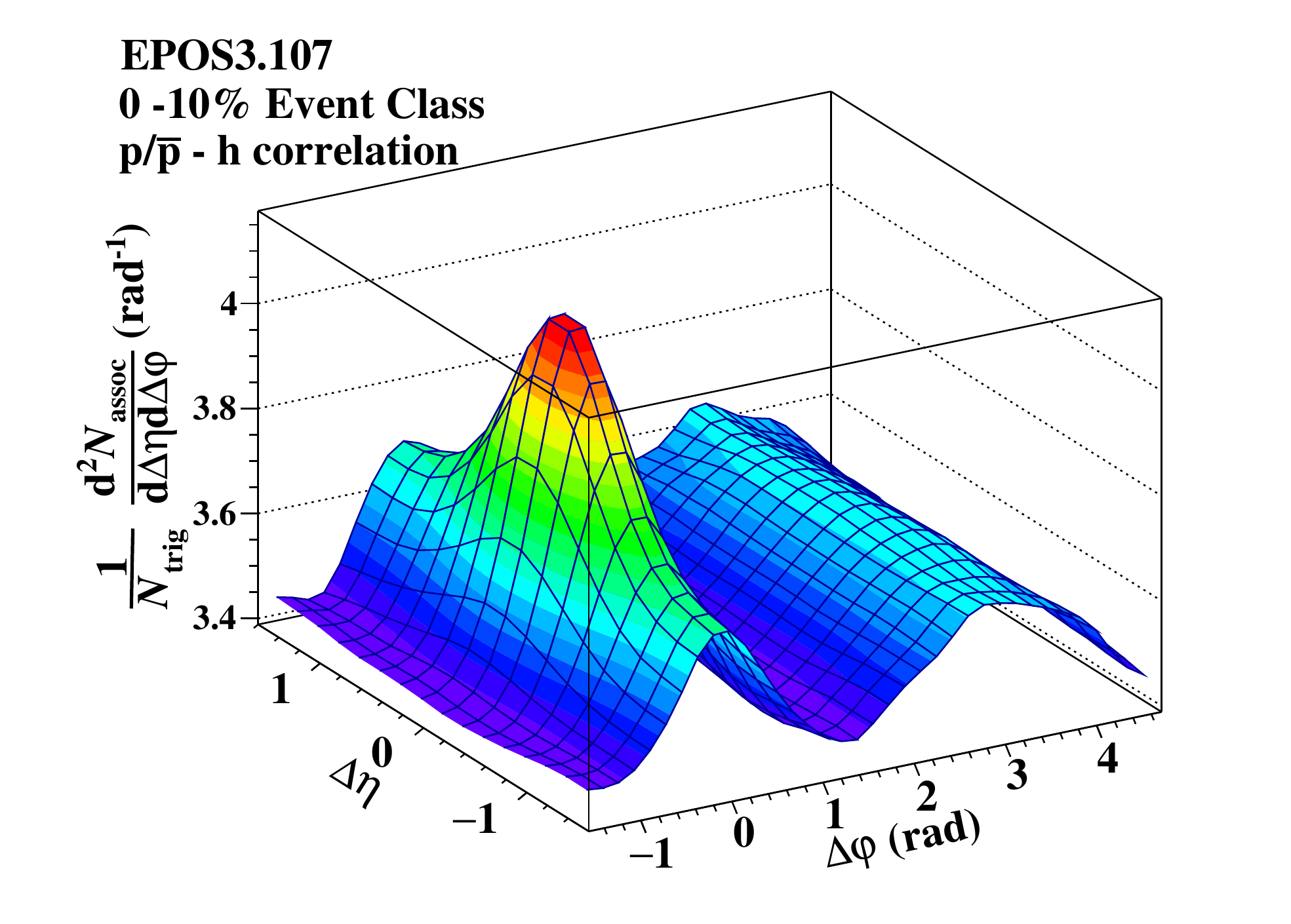}
\caption{[Color online] Two particle $\Delta\eta$-$\Delta\phi$ correlation function in 0-10 $\%$ event class 
of p-Pb collisions at $\sqrt{s_{NN} }$ = 5.02 TeV from EPOS with pion (left) and proton (right) as trigger particles.}
\label{corr-p-pi}
\end{figure}

This analysis concentrates only on the near side ($|\Delta\phi|$  $< \pi/2$) of the correlation function.
The particles from jet fragmentation are expected to be confined in a small angular region. To isolate  the near side jet-like correlation, we need to subtract
the modulation in $\Delta\phi$ arising out of the correlation with the event plane as represented by 
$v_2$, $v_3$ or higher harmonics. In this analysis the flow modulated background is estimated from large $|\Delta\eta|$ ($|\Delta\eta|$ $\geq$ 1.1) and subtracted from the near side jet peak ($|\Delta\eta| <$ 1.1) as it is done in  \cite {MPI_pPb}.

\begin{figure}
\begin{center}
\includegraphics[height=7.0 cm, width=8.0 cm]{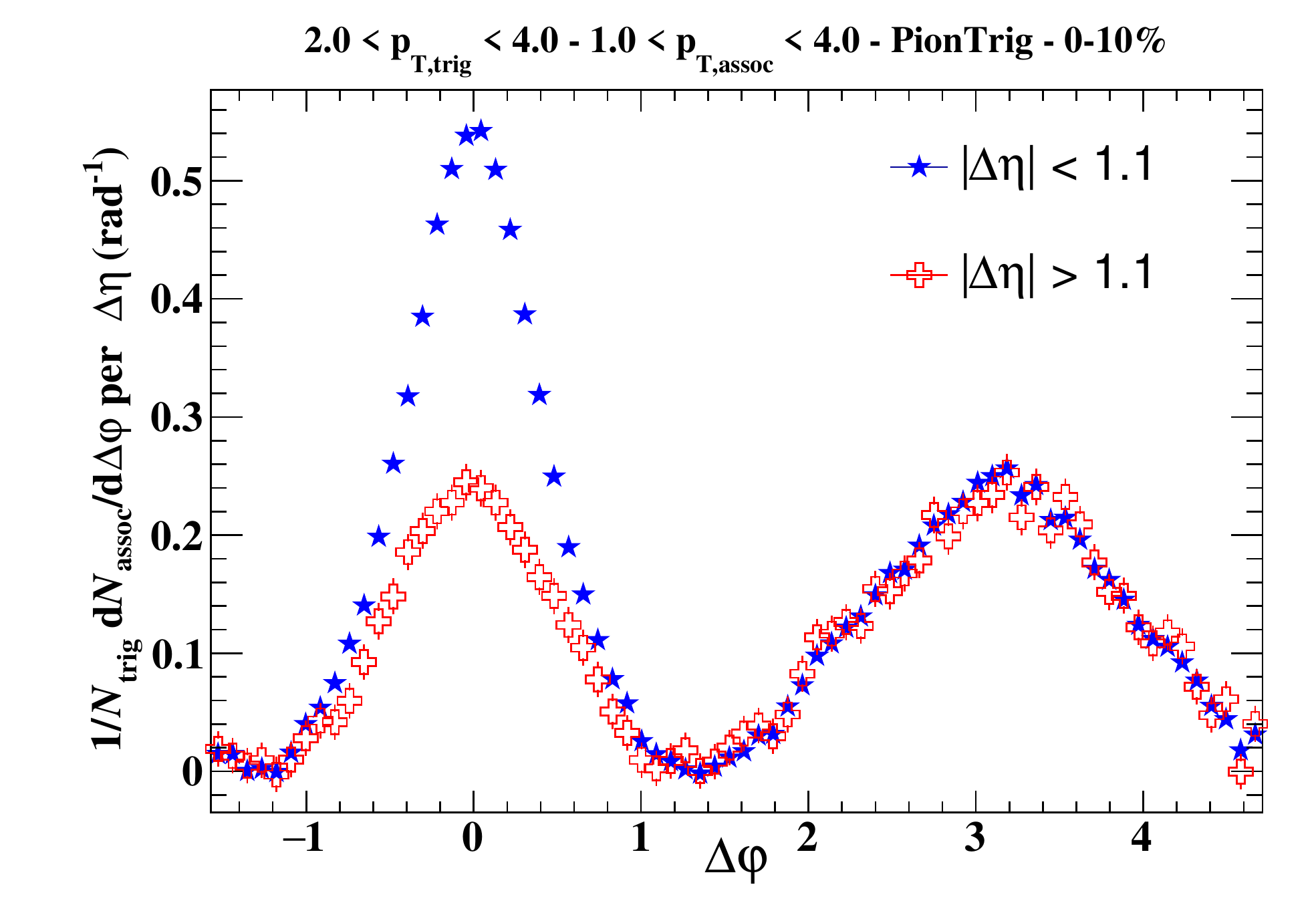}
\caption{[Color online] The  $\Delta\phi$ projected correlation function for two $\Delta\eta$ regions referred
to as jet (blue) and bulk (red).}
\label{proj-two-deleta}
\end{center}
\end{figure}

\begin{figure}
\begin{center}
\includegraphics[height=7.0 cm, width=8.0 cm]{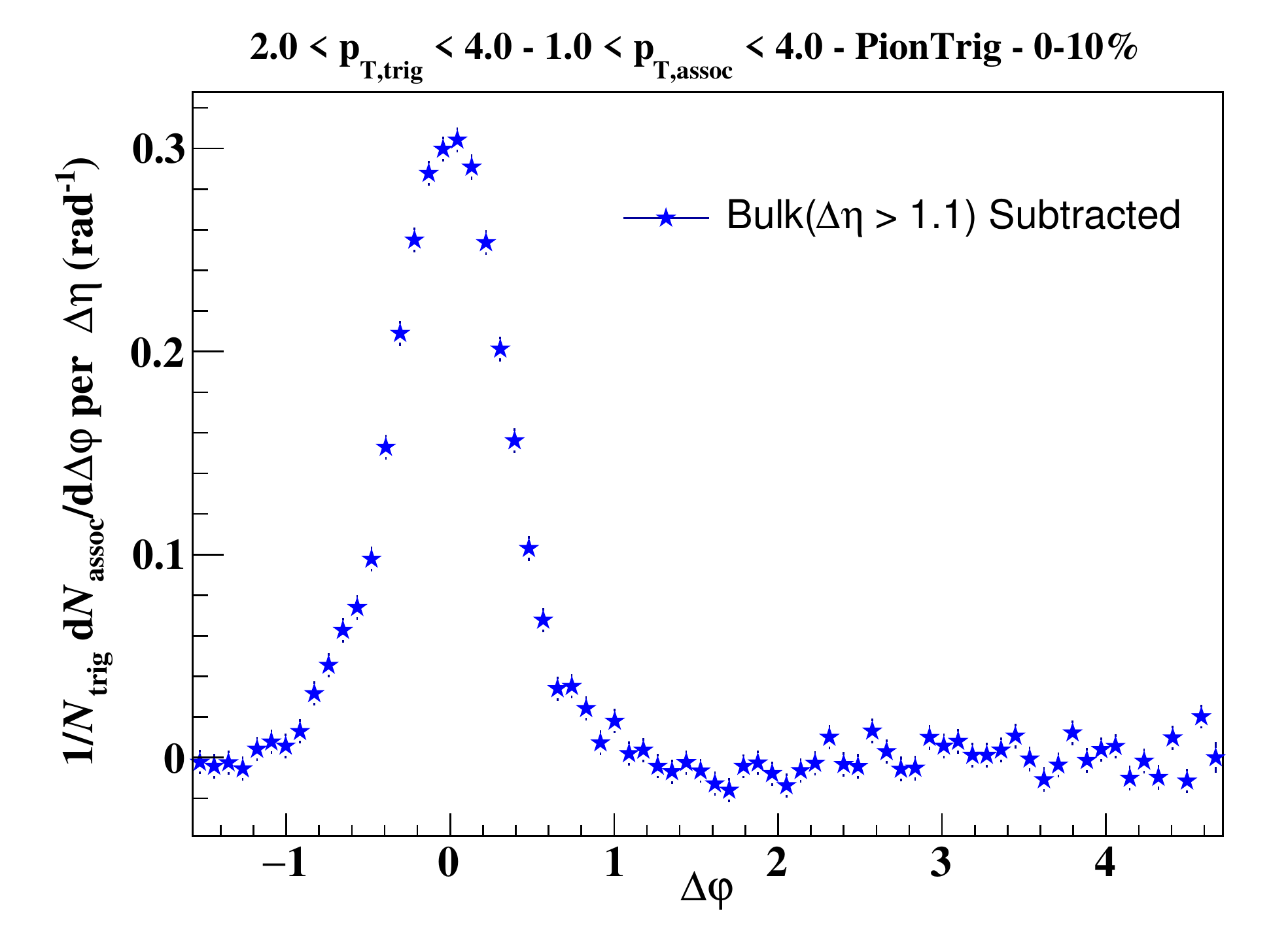}
\caption{[Color online] The  $\Delta\phi$ projected correlation function after bulk subtraction as discussed in the text.}
\label{proj-bkg-subtracted}
\end{center}
\end{figure}

The $\Delta\phi$ projected correlation functions for regions $|\Delta\eta| <$ 1.1 (jet) and $ |\Delta\eta| > $1.1 (bulk) are shown in Fig.~\ref{proj-two-deleta}. 
The background subtracted $\Delta\phi$ projected correlation function is shown in Fig.~\ref{proj-bkg-subtracted}. After bulk subtraction the event averaged near side jet like per trigger yield is calculated integrating the
$\Delta\phi$ projection in the range  $|\Delta\phi|$  $< \pi/2$.

\section{Results and Discussion}
In this analysis, two particle angular correlation measurements are performed by selecting the trigger particles from an intermediate $p_{T}$ range (2.0  $<p{_T}<$ 4.0 GeV/$\it{c}$), whereas, the associated particles have $p_{T}$ in the range of 1.0  $<p{_T}<$ 4.0 GeV/$\it{c}$.
The intriguing feature of particle production at intermediate transverse momentum is the contribution from
both hard and soft processes. In the ambit of EPOS model, particles originating from  {\bf corona} are referred to as hard particles and those from the {\bf core} are soft.\\ 
As per the EPOS model,
the two-dimensional correlation structure as obtained from the angular correlation measurements in p-Pb collisions at $\sqrt{s_{NN} }$ = 5.02 TeV reveal a near side jet peak over the so-called "ridge" structure extended over a large $|\Delta\eta|$ as shown in Fig 2. At such a collision energy, the origin of ridge structures in small collision systems was found to be reasonably described by the 3+1D hydro-calculations as implemented in the EPOS model \cite {epos_massordering_flow_pPb}. Particles originating from the core (soft particles) undergoes hydrodynamical evolution and expected not to have correlated partners beyond the "ridge" or flow like
correlations. Thus bulk subtracted near side jet
peak is dominated by the hard triggered (origin: corona) correlation. The bulk is estimated from the large $|\Delta\eta|$ ($|\Delta\eta|$ $\geq$ 1.1) and subtracted from the correlation function at small $|\Delta\eta|$ ($|\Delta\eta|$ $<$ 1.1). 
The soft triggers having no small angle correlated hadrons in the bulk subtracted near side jet peak is expected to create a dilution in the per trigger yield. \\ 

\begin{figure}[htb!]
\includegraphics[scale=0.37]{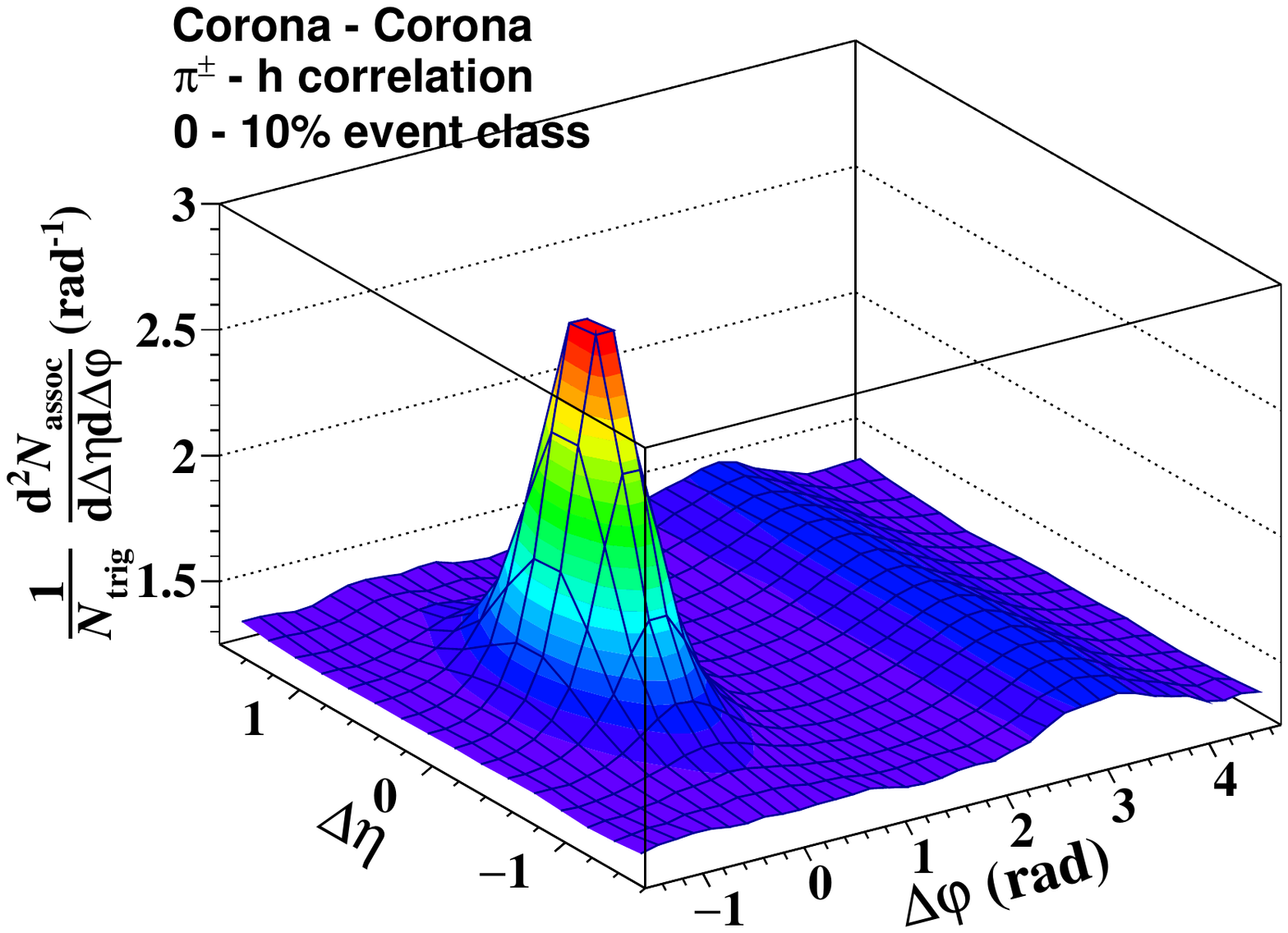}
\includegraphics[scale=0.36]{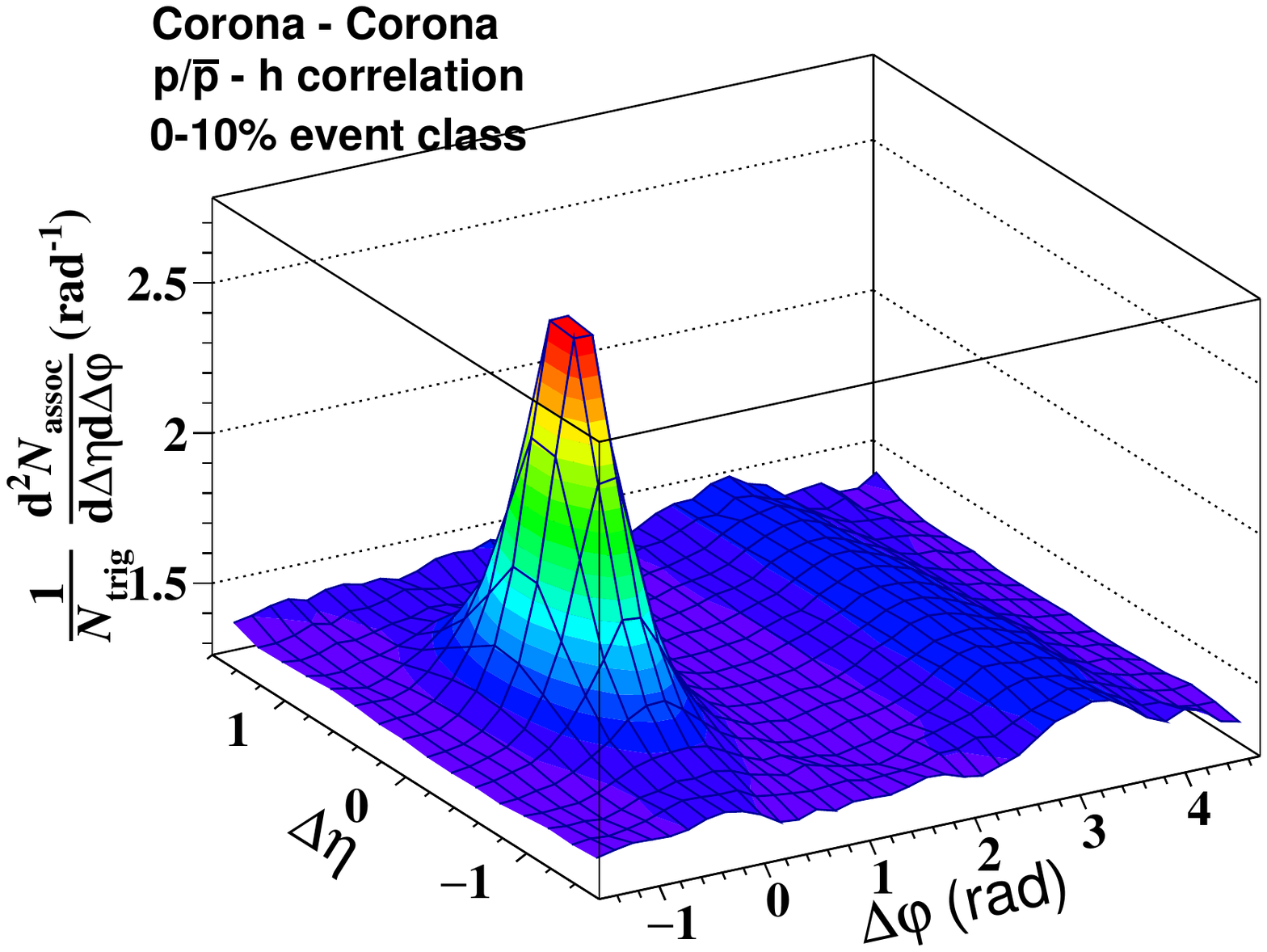}
\caption{[Color online] Two particle $\Delta\eta$-$\Delta\phi$ correlation function in 0-10 $\%$ event class 
of p-Pb collisions at $\sqrt{s_{NN} }$ = 5.02 TeV in case of corona-corona correlation from EPOS with pion (left) and proton (right) as trigger particles.}
\label{corr-p-pi}
\end{figure}

To investigate in detail the role of core or corona particles in the trigger dilution, the multiplicity
evolution of the hard triggered correlation (i.e. corona - corona correlation) has
been studied. Particles, stemming out-of the corona only are chosen to
construct 2D angular correlations (Fig. 5) that show a prominent near-side jet peak
 over a nearly flat baseline. The multiplicity dependence of the near side jet yield (baseline subtracted) in corona - corona correlations are shown in Fig 6. Both the pion and proton triggered yields increase with multiplicity but the rate of increase has no trigger species dependence. The ratio of the proton to pion triggered yield, shown in Fig 7, remains almost constant
as a function of multiplicity, showing no trigger dilution.

Now, in the calculation considering all (both core and corona) particles, there is a near
side jet peak over the so-called “ridge” structure extended over a large $|\Delta\eta|$
as already shown in Fig 2. As shown in Fig 4, the bulk subtracted near side jet peak is formed by
the hard triggered (corona - corona) correlation. But the correlation function
is normalized by both soft and hard triggers and the soft triggers without small
angle correlated hadrons in the bulk subtracted near side jet peak are expected to generate the trigger dilution.

\begin{figure*}[htb!]
\begin{center}
\includegraphics[scale=0.39]{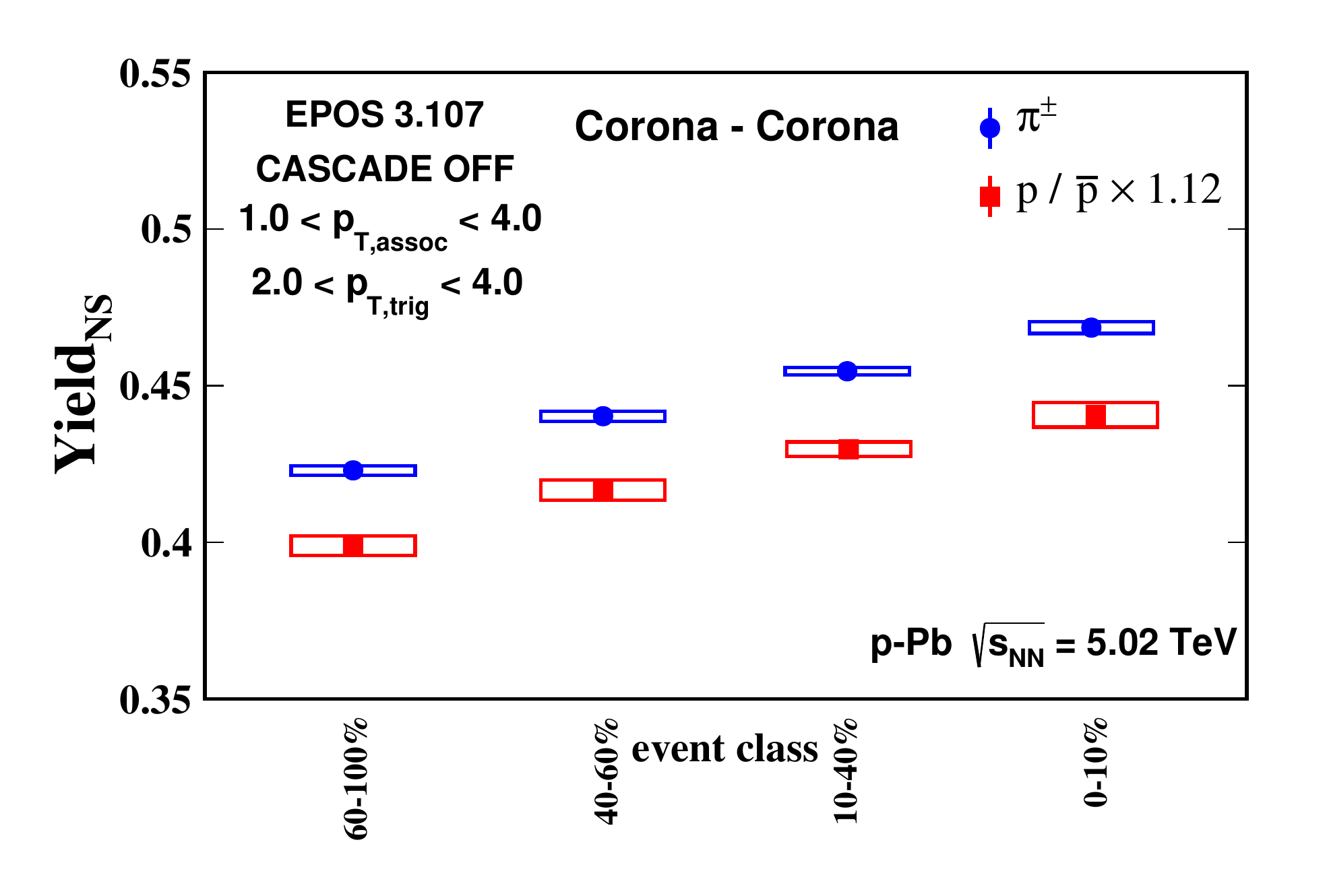}
\caption{[Color online] Multiplicity dependence of the near-side jet like per trigger yield associated with pion and proton triggers in case of corona - corona correlation in p-Pb collisions at $\sqrt{s}$ = 5.02 TeV from EPOS 3.107. Near-side jet yield has been calculated turning
hadronic cascade off.}
\end{center}
\label{individual_pr_pi_yield}
\end{figure*}

\begin{figure}[htb!]
\begin{center}
\includegraphics[scale=0.40]{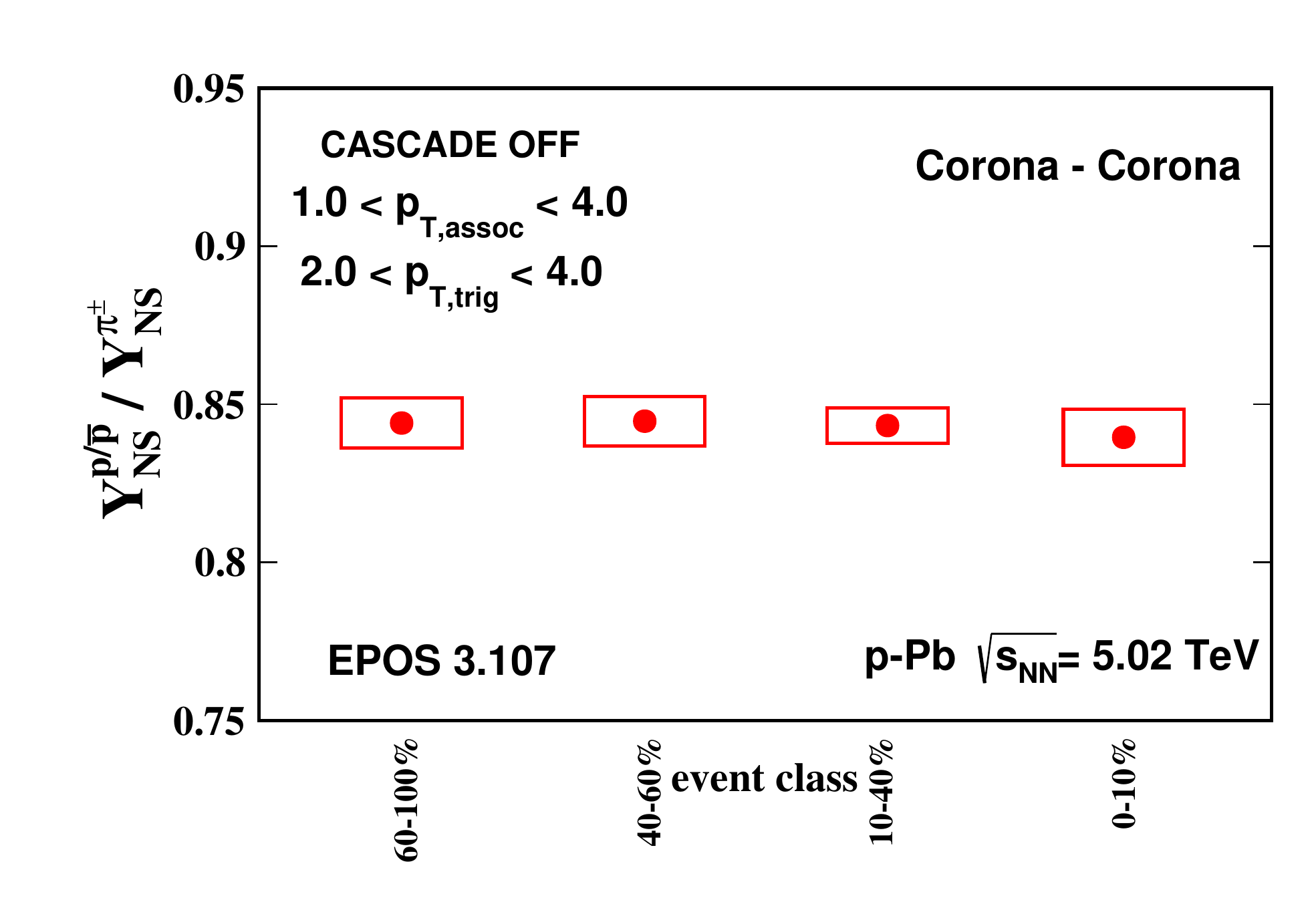}
\caption{[Color online] Multiplicity dependence of the ratio of the near-side jet like per trigger yield associated with proton and pion triggers in case of corona - corona correlation in p-Pb collisions at $\sqrt{s}$ = 5.02 TeV from EPOS 3.107. This ratio has been calculated turning
hadronic cascade off.}
\end{center}
\label{Yield_ratio_Pr_Pi}
\end{figure}

As we move from the lowest to highest multiplicity, the proportion of soft triggers increase and thus create a larger dilution in the higher multiplicity classes. The rate of dilution is associated with the rate of increase in the proportion of soft triggers which has a species dependence. In a hydro model like EPOS, radial flow pushes more protons than pions from lower to higher $p_{T}$ (2.0  $<p{_T}< $4.0 GeV/$\it{c}$) creating a baryon to meson enhancement as shown in Fig 1. So, a larger rate of dilution is expected in proton triggered correlation compared to the pion triggered case.\\
In Fig. 8 the multiplicity dependence of the bulk subtracted near side jet yield (per trigger) is shown. The pion triggered yield shows almost no variation with multiplicity whereas the proton triggered yield decreases gradually with increase in multiplicity. Comparison with Fig. 6 shows that both pion and proton triggered yields dilute with multiplicity but the proton triggered yield has a larger rate of dilution. The different trend in the multiplicity dependence of the pion and proton triggered yields can be attributed to the larger rate of increase of soft proton triggers (pushed by radial flow from lower to higher $p_{T}$) with multiplicity.  In Fig. 9 the multiplicity dependence of the ratio of proton to pion triggered yield is shown. A similar dilution pattern is observed from the lowest to highest multiplicity event classes with both hadronic cascade on and off - indicating hydrodynamics as a source of the trigger dilution observed here. The results from corona-corona only and all particles as shown in Fig. 7 and Fig. 9 conclude that the expected dilution comes only from the core.
\begin{figure*}[htb!]
\begin{center}
\includegraphics[scale=0.39]{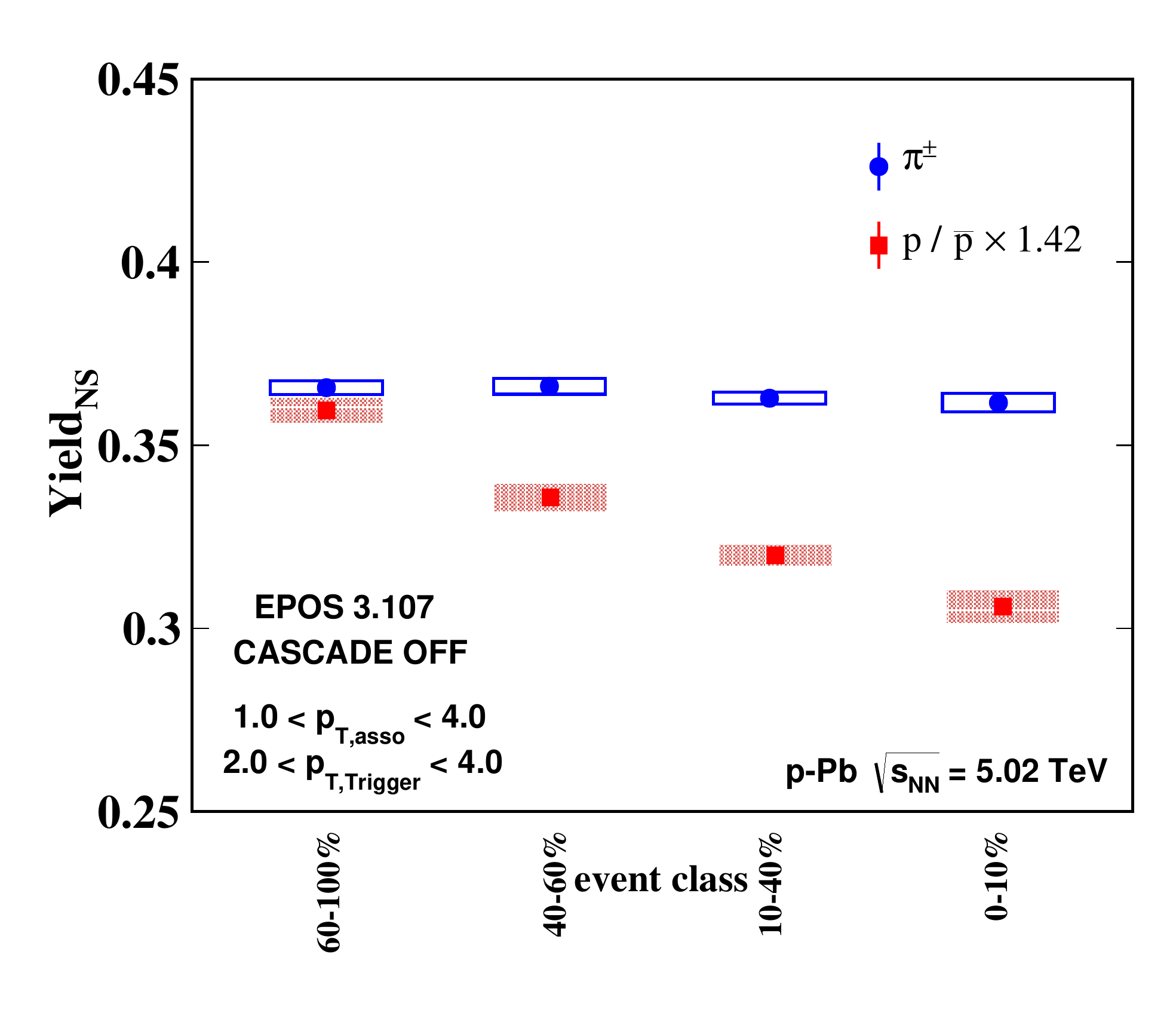}
\includegraphics[scale=0.41]{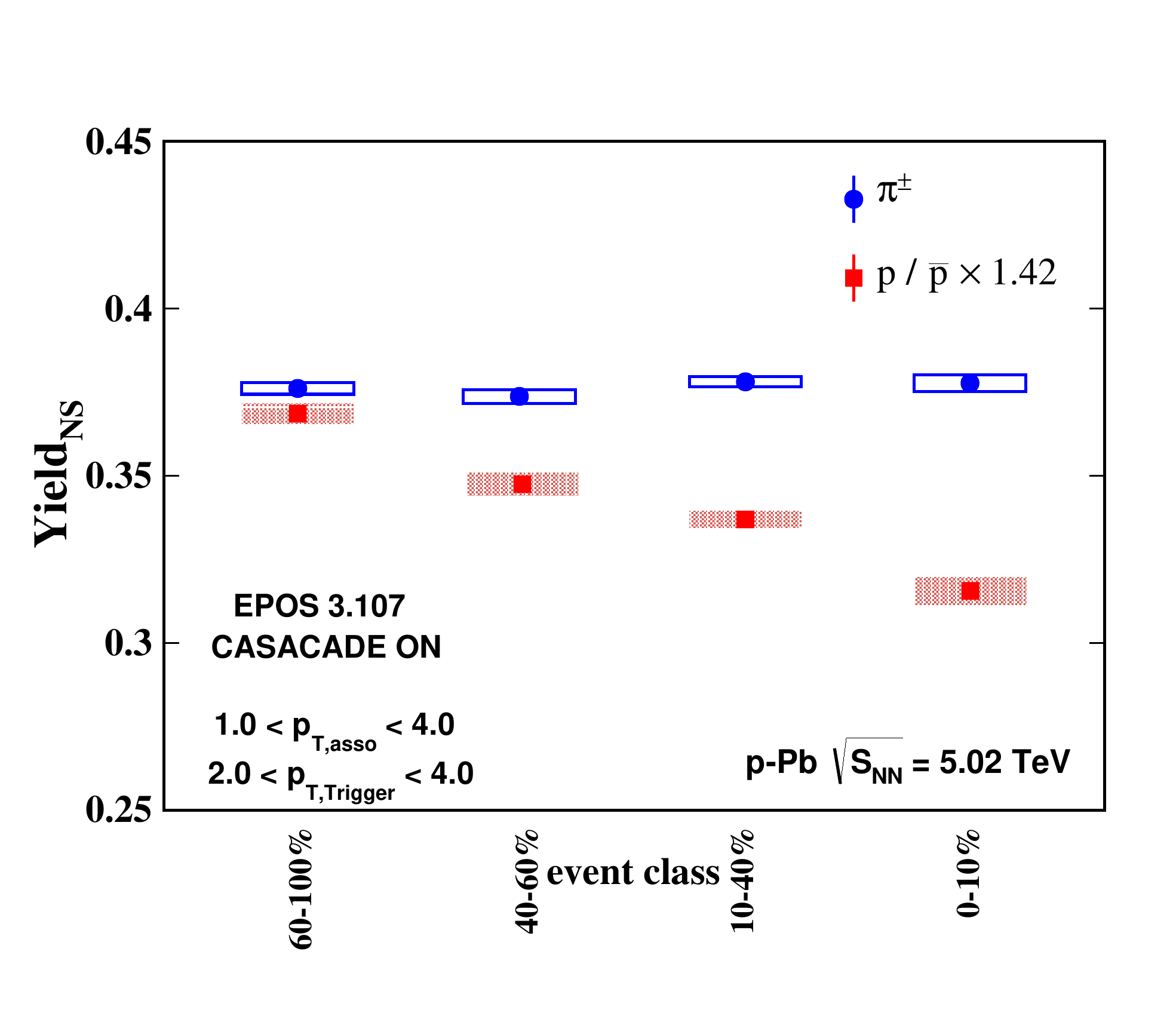}
\caption{[Color online] Multiplicity dependence of the near-side jet like per trigger yield (bulk subtracted) associated with pion and proton triggers (particles from both core and corona are considered)
in p-Pb collisions at $\sqrt{s}$ = 5.02 TeV from EPOS 3.107. Near-side jet yield has been calculated turning
hadronic cascade off (top) and on (bottom).}
\end{center}
\label{individual_pr_pi_yield}
\end{figure*}

\begin{figure}[htb!]
\begin{center}
\includegraphics[scale=0.40]{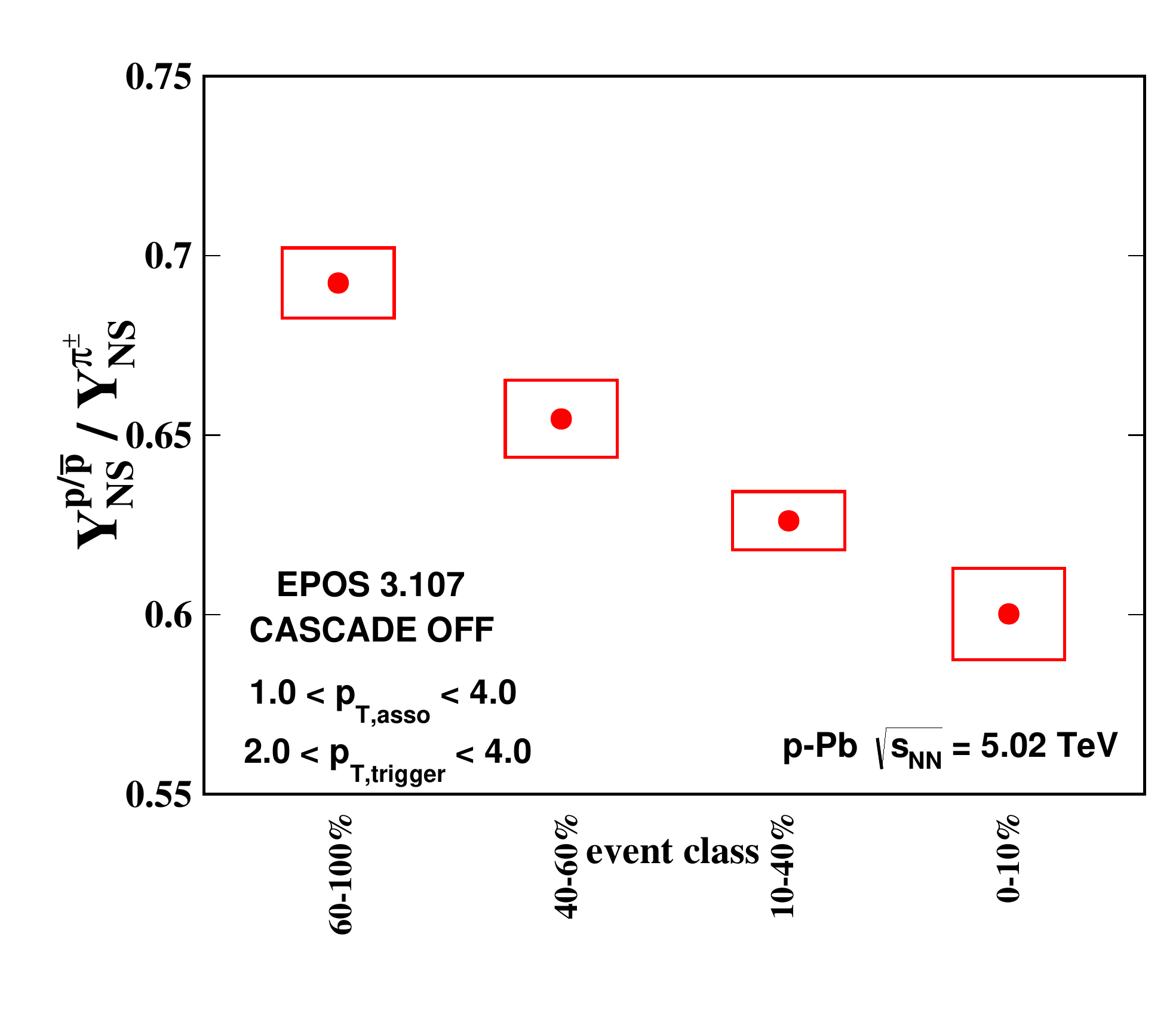}
\includegraphics[scale=0.40]{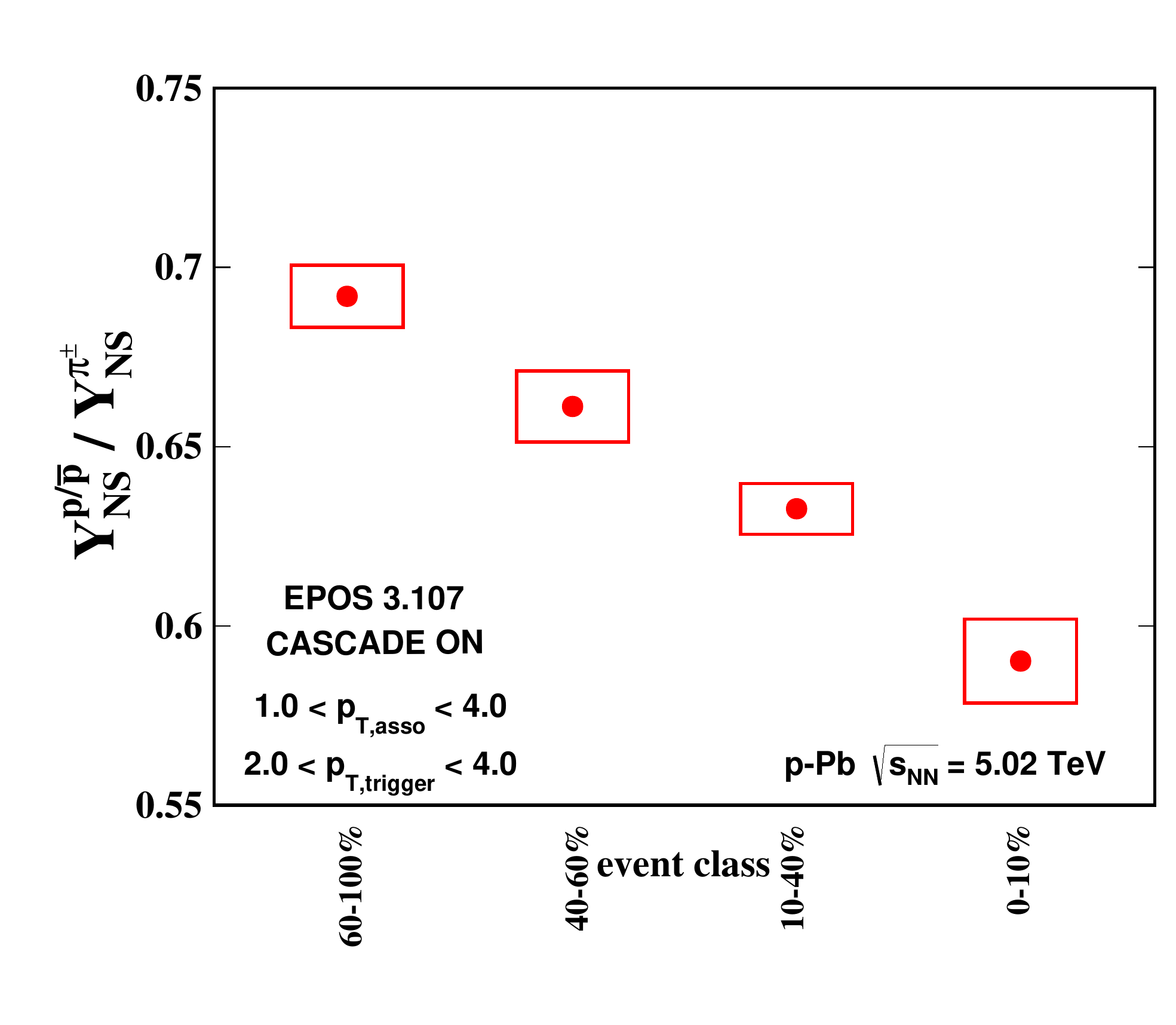}
\caption{[Color online] Multiplicity dependence of the ratio of the near-side jet like per trigger yield (bulk subtracted) associated with proton and pion triggers (particles from both core and corona are considered)
in p-Pb collisions at $\sqrt{s}$ = 5.02 TeV from EPOS 3.107. This ratio has been calculated turning
hadronic cascade off (top) and on (bottom).}
\end{center}
\label{Yield_ratio_Pr_Pi}
\end{figure}

A similar study has been performed by the PHENIX \cite {phenix_triggerdilution} and STAR collaborations \cite {star_triggerdilution} in min-bias pp, d-Au and Au-Au collisions at $\sqrt{s_{NN}} = $ 200 GeV. A hint of trigger dilution has been observed in the most central Au-Au collisions compared to the min-bias d-Au/ pp. The pion triggered yield increases in the most central Au-Au collisions compared to the min-bias d-Au whereas proton triggered yield remains almost unchanged or slightly reduced. The different trend in pion and proton triggered yields from min-bias d-Au to the most central Au-Au collisions has been argued to be a combined effect of competitive processes that include soft physics processes (coalescence model of hadronization and/or radial flow) and jet-medium interplay \cite {star_triggerdilution} \cite {phenix_triggerdilution}. A larger modification in the proton(non pion) triggered yield was expected as a larger fraction of proton triggers are predicted to be generated from gluon jets rather than quark jets compared to pion triggers \cite {star_triggerdilution}. Due to the difference in the colour charge, a larger medium induced energy loss by gluon jets is expected - resulting in higher jet like yields for proton(non pion) triggers \cite {star_triggerdilution}. However, lack of any such enhancement in data \cite {star_triggerdilution} has been attributed to the dominance of soft processes towards baryon production at intermediate $p_{T}$. It should be noted that in A-A collisions where jet quenching is prominent, the quenched energy is expected to manifest itself in terms of particles at low and intermediate $p_{T}$ possibly affecting both jet and bulk in a way that is yet to be understood  un-ambiguously. In small collision systems, like p-Pb, where medium induced modification of jets is observed to be less significant compared to that in the heavy ions \cite {RpPb} \cite {epos_radialflow_spectra_pPb}, a larger suppression in proton triggered yield(creating trigger dilution) can be dominantly associated with the presence of soft triggers at intermediate $p_{T}$ created by soft processes without significant contribution from the jet-medium interplay.\\
In this study, it has been shown that a model (EPOS) having hydrodynamical flow can generate trigger dilution in small collision systems (p-Pb) at LHC energy. Therefore hydrodynamics can be taken as an alternative to other explanation like coalescence model of hadronization which has been argued to be one of the possible reasons for trigger dilution in \cite {star_triggerdilution}  \cite {phenix_triggerdilution}. This observation puts a strong motivation for such a study by the LHC experiments as this observable may serve as an useful probe to investigate the presence of collective dynamics in small collision systems.
Further comparison with the data at LHC energy will be essential to constrain different models aiming to explain the observed signatures of collective behaviours in small collision systems.

\section*{Acknowledgements} 
We acknowledge fruitful discussions and suggestions from Dr. Klaus Werner and Dr. Federico Antinori at various stages of this work. We also thank Dr. Anders Garritt Knospe for helping us to set up initial framework for EPOS data generation. Thanks to VECC grid computing team for their 
constant effort to keep the facility running and helping in EPOS data generation.

\bibliography{mybibfile}

\end{document}